\documentclass[english]{IEEEtran}
\usepackage[T1]{fontenc}
\usepackage{textcomp}
\usepackage[latin9]{luainputenc}
\usepackage{array}
\usepackage{float}
\usepackage{booktabs}
\usepackage{multirow}
\usepackage{amsmath}
\usepackage{amssymb}
\usepackage{graphicx}

\makeatletter

\providecommand{\tabularnewline}{\\}
\floatstyle{ruled}
\newfloat{algorithm}{tbp}{loa}
\providecommand{\algorithmname}{Algorithm}
\floatname{algorithm}{\protect\algorithmname}

\usepackage[T1]{fontenc}
\usepackage[latin9]{inputenc}
\usepackage{array}
\usepackage{float}
\usepackage{booktabs}
\usepackage{multirow}
\usepackage{amsbsy}
\usepackage{amstext}
\usepackage{graphicx}

\usepackage{cite}
\usepackage{xcolor}
\usepackage{hyperref}
\usepackage{babel}

\usepackage{amsmath}
\usepackage{amsfonts}
\usepackage{amssymb}
\usepackage{amsthm}
\usepackage{tabularx}
\usepackage{mathrsfs} 

\usepackage{url}
\usepackage{hyperref}

\usepackage{stfloats}
\usepackage{float}
\usepackage{graphicx}
\usepackage{cite}
\usepackage{xcolor}
\usepackage{subfigure}

\makeatletter
\def\blfootnote{\xdef\@thefnmark{}\@footnotetext}
\makeatother
\begin{document}
\title{A Constant-Time Hardware Architecture for the CSIDH Key-Exchange Protocol}
\author{Sina Bagheri, Masoud~Kaveh, \IEEEmembership{Member, IEEE}, Francisco Hernando-Gallego, Diego Mart\'in, and Nuria Serrano
}
\IEEEspecialpapernotice{
\thanks{S. Bagheri is with the Department of Electrical and Computer Engineering,
University of Tehran, Tehran, Iran. (e-mail: $\rm sina.bagheri80@ut.ac.ir$)} 
\thanks{M. Kaveh is with the Department of Information and Communication Engineering, Aalto University, Espoo, Finland. (e-mail: $\rm masoud.kaveh@aalto.fi$)}
\thanks{F. Hernando-Gallego, D. Mart\'in, and N. Serrano are with the Department of Computer Science, Escuela de Ingenier\'ia Inform\'atica de Segovia, Universidad de Valladolid, Segovia, 40005, Spain (e-mail:$\rm \{fhernando, diego.martin.andres, nuria.serrano\}@uva.es$)}
}
\maketitle
\begin{abstract}
The commutative supersingular isogeny Diffie-Hellman (CSIDH) algorithm
is a promising post-quantum key exchange protocol, notable for its
exceptionally small key sizes, but hindered by computationally intensive
key generation. Furthermore, practical implementations must operate
in constant time to mitigate side-channel vulnerabilities, which presents
an additional performance challenge. This paper presents, to our knowledge,
the first comprehensive hardware study of CSIDH, establishing a performance
baseline with a unified architecture on both field-programmable gate
array (FPGA) and application-specific integrated circuit (ASIC) platforms.
The architecture features a top-level finite state machine (FSM) that
orchestrates a deeply pipelined arithmetic logic unit (ALU) to accelerate
the underlying 512-bit finite field operations. The ALU employs a
parallelized schoolbook multiplier, completing a 512$\times$512-bit multiplication
in 22 clock cycles and enabling a full Montgomery modular multiplication
in 87 cycles. The constant-time CSIDH-512 design requires $1.03\times10^{8}$
clock cycles per key generation. When implemented on a Xilinx Zynq
UltraScale+ FPGA, the architecture achieves a 200 MHz clock frequency,
corresponding to a 515 ms latency. For ASIC implementation in
a 180nm process, the design requires $1.065\times10^{8}$ clock cycles
and achieves a \textasciitilde 180 MHz frequency, resulting in a
key generation latency of 591 ms. By providing the first public hardware
performance metrics for CSIDH on both FPGA and ASIC platforms, this
work delivers a crucial benchmark for future isogeny-based post-quantum cryptography (PQC) accelerators.
\end{abstract}


\section{Introduction}

The advent of practical quantum computers threatens to undermine the security of widely used public-key schemes such as Rivest-Shamir-Adleman (RSA) and elliptic curve cryptography (ECC), accelerating
the search for post-quantum cryptography (PQC) that remains secure
even in the presence of quantum adversaries. In response to this emerging
threat, the National Institute of Standards and Technology (NIST)
initiated a global standardization process to identify and evaluate
cryptographic schemes suitable for widespread practical use. This
effort has driven the development of algorithms from several distinct
families, including lattice-based schemes like CRYSTALS-Kyber and
CRYSTALS-Dilithium, as well as isogeny-based cryptography \cite{Intro 1,Dilithium saeed,key-1,Kyber main,deleithum main}. As many
of these PQC proposals rely on complex mathematical operations, hardware
acceleration is generally expected to be critical for
achieving acceptable performance and for implementing side-channel
countermeasures in resource-constrained environments \cite{Intro 1}.

Among PQC families, isogeny-based cryptography has attracted attention
for its extremely compact parameters. These schemes exploit the hardness
of finding isogenies (structure-preserving maps) between super-singular
elliptic curves. Early work by Jao and De Feo (SIDH) \cite{intro2}
demonstrated a key-exchange based on supersingular isogenies. However,
SIDH requires an interactive (ephemeral) protocol and additional transformations
(such as SIKE\textquoteright s Fujisaki\textendash Okamoto wrapping)
to achieve chosen-ciphertext security. More recently, Castryck et
al. \cite{Main CSIDH Article} proposed CSIDH, a commutative supersingular
isogeny Diffie\textendash Hellman scheme, which realizes a commutative
group action on supersingular curves. This commutative property enables
two parties to each apply a secret (a \textquotedblleft class group\textquotedblright{}
element) to a common base curve in a non-interactive, static\textendash static
Diffie\textendash Hellman fashion. In other words, CSIDH supports
a one-round key exchange using fixed public keys, a feature that can
significantly simplify protocol design \cite{Main CSIDH Article}.

CSIDH offers several appealing strengths. Its public keys are extraordinarily
small: for example, the CSIDH-512 instantiation achieves \ensuremath{\sim}128-bit
security with only 64-byte public keys, making it one of the most
compact of all PQC candidates. This is far smaller than the kilobyte-sized
keys of lattice or code-based schemes. The underlying construction
also allows straightforward public-key validation: as noted by Castryck
et al. \cite{Main CSIDH Article}, CSIDH \textquotedblleft allows
public-key validation at very little cost\textquotedblright . In practice,
this means that public keys and shared secrets can be confirmed efficiently
without expensive transforms, ensuring robustness against invalid-curve
attacks. Moreover, by design CSIDH\textquoteright s commutative action
gives it the \textquotedblleft static\textendash static\textquotedblright{}
(non-interactive) key exchange property that many other PQ schemes
lack. These features make CSIDH attractive for applications requiring
low-bandwidth secure key agreement \cite{Main CSIDH Article,main constant time}.

\subsection{Related Works and Research Gaps}
Despite these advantages, CSIDH suffers from significant performance
drawbacks. The group action in CSIDH requires composing a large number
of supersingular isogenies of varying degrees, which is computationally
intensive. For instance, the original CSIDH reference implementation
(Intel Skylake, 3.5\,GHz) reports that a single group-action (applying
one party\textquoteright s secret) takes on the order of 40 milliseconds
\cite{Main CSIDH Article}. Constant-time implementations further
increase cost as dummy-isogeny techniques and other countermeasures
are necessary to thwart timing and fault attacks. As Campos et al.
\cite{main constant time} observe, CSIDH is \textquotedblleft inherently
difficult to implement in constant time\textquotedblright , and even
state-of-the-art dummy-free algorithms incur about a 2\texttimes{}
slowdown relative to variable-time code. In practice, this means that
CSIDH key exchanges can take seconds on embedded processors \cite{jalali work on constant time}.
In contrast, lattice-based key encapsulations (e.g. Kyber) or even
SIDH/SIKE implementations can complete in milliseconds. Thus, while
CSIDH\textquoteright s small keys are compelling, its runtime performance
remains the scheme\textquoteright s Achilles\textquoteright{} heel.
Recent works continue to optimize CSIDH (for example using elliptic-curve
point encodings and improved isogeny strategies), but the inherent
computational burden remains high \cite{optimize constant time2}.

In comparison to other PQ approaches, CSIDH exhibits unique trade-offs.
Lattice and code schemes have much larger public keys (often thousands
of bytes) but are faster to compute and have been widely accelerated
in software and hardware. Multivariate schemes can be fast but also
tend to have large signatures or public keys. In the isogeny realm,
SIDH/SIKE and CSIDH share the advantage of compact keys, but SIDH/SIKE
requires an interactive protocol and large transient public-key components
during the handshake. By contrast, CSIDH\textquoteright s static\textendash static,
commutative design simplifies protocols at the cost of slower math.
Moreover, unlike many lattice schemes, CSIDH\textquoteright s security
is based on well-studied hard problems in elliptic-curve theory (currently
believed to be resistant to both classical and quantum attacks), giving
it strong security appeals alongside its compactness \cite{Main CSIDH Article}. 

Crucially, almost all reported CSIDH implementations to date are in
software. Castryck et al. published a C/Sage reference implementation
of CSIDH \cite{Main CSIDH Article}. Meyer et al. demonstrated an
efficient constant-time C implementation using dummy operations and
mapping tricks \cite{optimize constant time2}, and Campos et al.
evaluated CSIDH on a Cortex-M4 microcontroller, analyzing fault injection
attacks \cite{main constant time}. Jalali et al. focused on 64-bit
high-performance ARM Cortex-A72 cores, presenting an implementation
where the constant-time group action takes 12 seconds \cite{jalali work on constant time}. However,
to our knowledge, no FPGA or ASIC implementation of CSIDH has appeared
in the literature. This is a notable gap, especially since isogeny
schemes like SIDH/SIKE have seen multiple hardware realizations achieving
dramatic speed-ups. Without dedicated hardware support, CSIDH\textquoteright s
slowness may preclude its adoption in practice. It is therefore essential
to explore hardware acceleration to make CSIDH practical and efficient
for real-world use.

\subsection{Paper Contributions}
In this paper, we aim to fill this gap by presenting a unified hardware
architecture for CSIDH\footnote{The complete Verilog RTL source code and testbenches for this project
are open-source and available on GitHub at: \href{https://github.com/sina1777/CSIDH}{https://github.com/sina1777/CSIDH}}. The main contributions in this paper can be summarized as follows:
\begin{itemize}
    \item We present, to the best of our knowledge, the first comprehensive hardware implementation of the CSIDH key-exchange protocol, targeting both FPGA and ASIC platforms. Our architecture accelerates the CSIDH group action by integrating fast, deeply pipelined Montgomery modular arithmetic units with a highly optimized finite state machine (FSM) that maximizes resource utilization and minimizes control overhead.
    \item We design a unified hardware framework capable of supporting both the 512-bit and 1024-bit CSIDH parameter sets. The architecture incorporates several novel microarchitectural optimizations, including a parallelized 512$\times$512-bit schoolbook multiplier, a pipelined carry-select adder, constant-time isogeny evaluation with integrated dummy operations, and an arithmetic logic unit (ALU) masking strategy to mitigate timing and power side-channel attacks.
    \item We provide a rigorous performance evaluation on two distinct hardware targets: a Xilinx Zynq UltraScale+ FPGA and a 180\,nm ASIC implementation using the SMIC process. Our measurements include area utilization, maximum operating frequency, latency, and throughput, establishing a realistic performance baseline for future CSIDH accelerators.
    \item Experimental results show that the proposed hardware significantly outperforms embedded software implementations, reducing constant-time CSIDH-512 key generation to 515\,ms on FPGA and 591\,ms on ASIC, and achieving multi-fold speedups compared to state-of-the-art microcontroller-based designs. These results demonstrate that the proposed accelerator makes CSIDH a more practical and deployable post-quantum key-exchange scheme for bandwidth- and resource-constrained environments.
\end{itemize}

\subsection{Paper Organization}
The remainder of this paper is organized as follows. Section~\ref{sec-pre} presents a detailed review of the CSIDH protocol. Section~\ref{sec-scheme} introduces the proposed hardware architecture and the implementation strategies adopted. In Section~\ref{sec-sec}, we provide a comprehensive security analysis of the proposed accelerator. Section~\ref{sec-results} reports the implementation results and offers an in-depth performance evaluation. Finally, Section~\ref{sec-conclusion} concludes the paper and discusses potential directions for future research.

\section{PRELIMINARIES} \label{sec-pre}
In this section, we provide an overview of the CSIDH algorithm. For a more comprehensive treatment of its underlying principles, the reader is referred to \cite{Main CSIDH Article}. The algorithmic steps are also summarized in Algorithm~\ref{algo1}.

\subsection{CSIDH}
At its heart, CSIDH works over a prime field $\digamma_{p}$ where
$p$ is chosen in special form $p=4l_{1}l_{2}\cdots l_{n}-1$ with
each $l_{i}$ are small odd primes. This ensures that every supersingular
elliptic curve over $\digamma_{p}$ have points of order $l_{i}$,
which allow us to efficiently build isogenies of degree $l_{i}$.
Each curve is represented in Montgomery form $E_{A}:y^{2}=x^{3}+Ax^{2}+x$
parametrized by $A\in F_{p}$. In particular, one fixes the base curve
$E_{0}:y^{2}=x^{3}+x$, whose endomorphism ring is $Z[\pi]$, and
works within the commutative class group $C1(Z[\pi])$ acting on the
set of all such supersingular curves \cite{Main CSIDH Article}. 

Each user\textquoteright s private key is an exponent vector $e=(e_{1},\cdots,e_{n})$
with $e_{i}\in[-m,m]$. This vector corresponds to the class group
element $[a]=[l_{1}^{e_{1}}l_{2}^{e_{2}}\cdots l_{n}^{e_{n}}]\in C1(Z[\pi])$.
To generate a public key, one applies the corresponding chain of isogenies
to the base curve $E_{0}$, yielding a new Montgomery coefficient
$A\in\digamma_{p}$; this coefficient alone serves as the public key.
A key exchange between Alice $(e_{A})$ and Bob $(e_{B})$ then proceeds
exactly like classical Diffie\textendash Hellman:
\begin{enumerate}
\item \textbf{Publish: }Alice computes $E_{A}=[a]E_{0}$ and publishes its
coefficient $A$; Bob similarly obtains B. 
\item \textbf{Compute: }Alice applies her secret $[a]$ to Bob's $E_{B}$,
and Bob applies $[b]$ to Alice's $E_{A}$.
\item \textbf{Agree: }Commutativity in $C1(Z[\pi])$ guarantees
$[a]([b]E_{0})=[b]([a]E_{0})$, so both arrive at the same final curve
$E_{AB}$. The shared secret is then extracted from its Montgomery
coefficient $S\in\digamma_{p}$ \cite{Main CSIDH Article}.
\end{enumerate}
\begin{algorithm}[t]
$\textbf{Input 1:}$ Input public key $A\in F_{p}$ 

$\textbf{Input 2:}$ Private key $e=(e_{1},e_{2},\cdots,e_{n})$ ,
$e_{i}\in[5:-5]$

$\textbf{Output:}$ Output public key $A'\in F_{p}$, Boolean success

\textemdash\textemdash\textemdash\textemdash\textemdash\textemdash\textemdash\textemdash\textemdash\textemdash\textemdash\textemdash\textemdash\textemdash\textemdash\textemdash\textemdash\textemdash\textemdash\textemdash\textemdash\textemdash\textemdash\textemdash\textemdash{}

1: \textbf{if }not validate\_basic($A$) \textbf{then}

2: \textbf{\quad{}$A'\leftarrow$}random element in $\digamma_{p}$

3: \textbf{\quad{}return }$(A',flase)$

4: \textbf{end if}

5: $A_{proj}\leftarrow(A,1)$

6: \textbf{for }twist in $[false,true]$ \textbf{do}

7:\textbf{\quad{}while} true \textbf{do}

8:\textbf{\quad{}\quad{}}batch $\leftarrow$ select up to 16 primes
i where ($e[i]>0\land\sim$

\textbf{\quad{}\quad{}\quad{}$twist)\lor(e[i]<0\land twist)$}

9:\textbf{\quad{}\quad{}if }batch is empty \textbf{then}

10:\textbf{\quad{}\quad{}\quad{}if }twist \textbf{then}

11:\textbf{\quad{}\quad{}\quad{}\quad{}break\,}

12:\textbf{\quad{}\quad{}\quad{}else}

13:\textbf{\quad{}\quad{}\quad{}\quad{}$twist\leftarrow true$}

14:\textbf{\quad{}\quad{}\quad{}\quad{}continue}

15:\textbf{\quad{}\quad{}\quad{}end if}

16:\textbf{\quad{}\quad{}endif}

17:\textbf{\quad{}\quad{} $k\leftarrow4\ast\prod_{j\notin batch}l_{j}$}

18:\textbf{\quad{}\quad{}$P\leftarrow$ }generate\_elligator\_point

19:\textbf{\quad{}\quad{}$P\leftarrow[k]P$}

20:\textbf{\quad{}\quad{}for }i in batch descending order \textbf{do}

21:\textbf{\quad{}\quad{}\quad{}}$cof\leftarrow\prod_{j\in batch,j<i}l_{j}$

22:\textbf{\quad{}\quad{}\quad{}$K\leftarrow[cof]P$}

23:\textbf{\quad{}\quad{}\quad{}if }is\_infinity$(K)$ \textbf{then}

24:\textbf{\quad{}\quad{}\quad{}\quad{}continue }

25:\textbf{\quad{}\quad{}\quad{}end if}

26:\textbf{\quad{}\quad{}\quad{}$xISOG(A_{proj},P,K,l_{i})$}

27:\textbf{\quad{}\quad{}\quad{}if not }$twist$

28:\textbf{\quad{}\quad{}\quad{}\quad{}$e[i]\leftarrow e[i]-1$}

29:\textbf{\quad{}\quad{}\quad{}else }

30:\textbf{\quad{}\quad{}\quad{}\quad{}$e[i]\leftarrow e[i]+1$}

31:\textbf{\quad{}\quad{}end for}

32:\textbf{\quad{}\quad{}}generate new random point

33:\textbf{\quad{}end while}

34:\textbf{ end for}

35:\textbf{ if }validate ($A_{proj}$) then

36:\textbf{\quad{}$A'\leftarrow A_{proj}.x/A_{proj}.z$}

37:\textbf{\quad{}return $(A',true)$}

38: \textbf{else}

39:\textbf{\quad{}$A'\leftarrow$}random element in\textbf{ $\digamma_{p}$}

40:\textbf{\quad{}return $(A',false)$}

\caption{CSIDH Public Key Genertion \cite{Main CSIDH Article}}
\label{algo1}
\raggedright{}\vspace{1mm}
\end{algorithm}

\subsection{Isogenies}

The core operation in CSIDH is to compute isogenies. Costello and Hisil
\cite{isogeny optimiztion} presented a simple and compact mathematical
algorithm for computing an isogeny $E\rightarrow E\prime$ of odd
degree $\ell=2d+1$. These formulas require a kernel point $K$, which
is a point of order $\ell$. Let $K\in E$ be a kernel point and $(X_{i}:Z_{i})$
is the projective coordinates of the point $[i]K$. With this, the
new point under isogeny can be calculated using the following formula
\cite{isogeny optimiztion,main constant time}:
\begin{multline}
\varphi:(X:Z)\mapsto\\
(X\left(\prod_{i=1}^{d}(X-Z)(X_{i}+Z_{i})+(X+Z)(X_{i}-Z_{i})\right)^{2}:\\
Z\left(\prod_{i=1}^{d}(X-Z)(X_{i}+Z_{i})-(X+Z)(X_{i}-Z_{i})\right)^{2})
\end{multline}
Furthermore, Meyer and Reith presented a more efficient method for
calculating the curve parameters, detailed as follows. Consider the
$A'=(A_{x}':A_{z}')$ the projective coordinate of new curve parmeter
\cite{Faster isogeny}:
\begin{multline}
(A_{x}':A_{z}')=(2\left((A+2)^{l}*\pi_{+}^{8}+(A-2)^{l}*\pi_{-}^{8}\right)^{2}:\\
\left((A+2)^{l}*\pi_{+}^{8}-(A-2)^{l}*\pi_{-}^{8}\right)),
\end{multline}
where $\pi_{+}=\prod_{i=1}^{d}(X_{i}+Z_{i})$ and $\pi_{-}=\prod_{i=1}^{d}(X_{i}-Z_{i})$.

Finally, note that CSIDH needs to apply many such isogenies (one for
each prime $l_{i}$ used in the secret). In implementations, intermediate
points are re-used across these steps whenever possible to avoid repeated
scalar multiplications. For example, once the kernel generator $Q=[(P+1)/l_{i}]P$
is found, the same projective point P can be carried forward through
the isogeny: after evaluating $\varphi:E_{A}\rightarrow E_{A'}$ one
sets $P\leftarrow\varphi(P)$ to obtain a point on the new curve.
This way, the next step's isogeny can begin from an already-scaled
point rather than a fresh random generator. Such reuse of intermediate
points, combined with batching of scalar multiplies and careful projective
arithmetic, yields the efficient isogeny walks that make CSIDH practical
\cite{Faster isogeny,faster isogeny2}.

\section{Proposed Hardware Implementation } \label{sec-scheme}

\begin{figure}
\includegraphics[bb=240bp 123bp 767bp 510bp,clip,width=1\columnwidth]{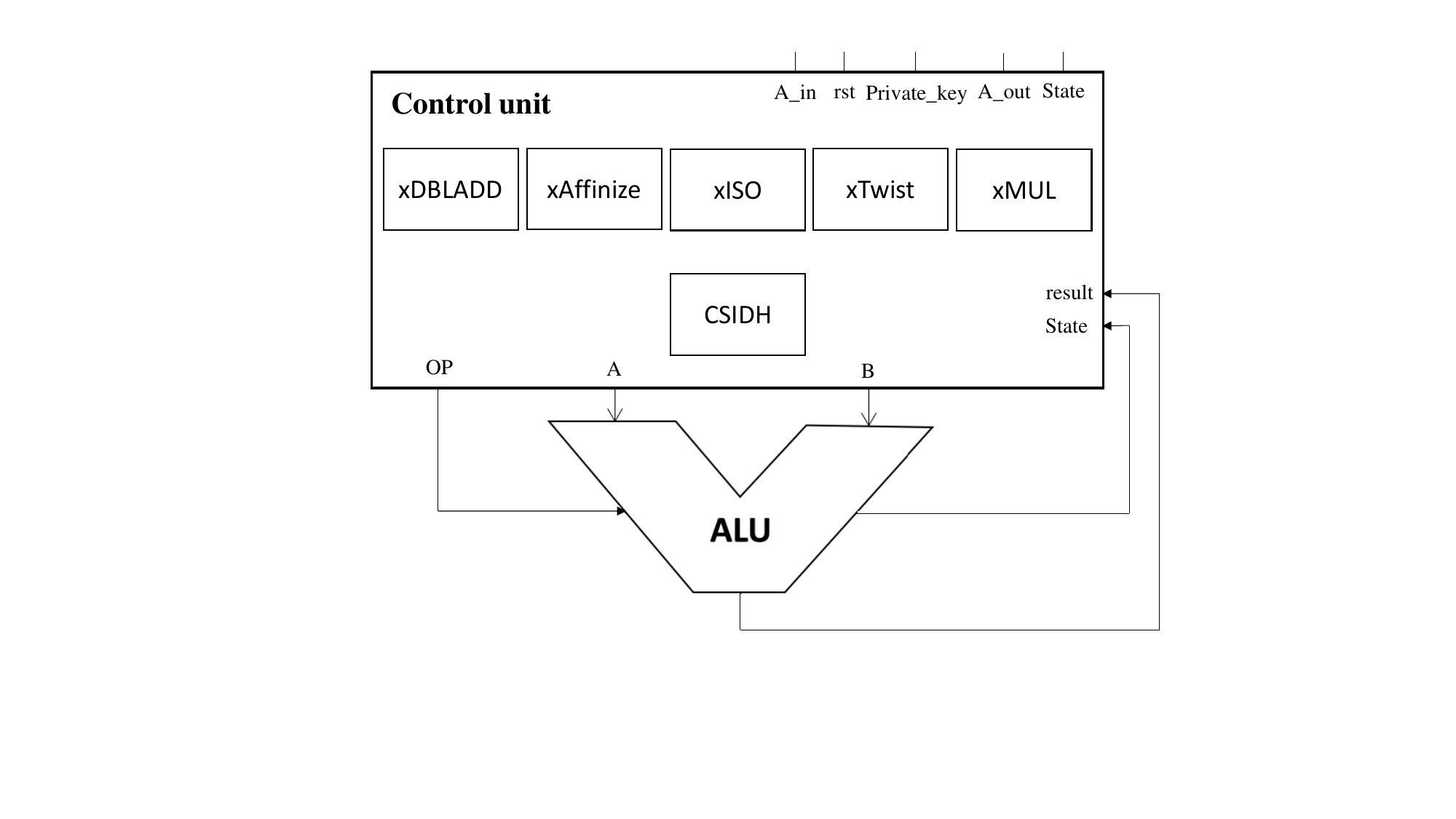}\caption{High-level architecture of the proposed CSIDH accelerator. The design
consists of a central Control Unit, which orchestrates the key generation
process, and a shared ALU that executes the required 512-bit finite
field operations.}
\label{fig:arch}
\end{figure}

This section details our proposed unified CSIDH hardware architecture,
designed for efficient implementation on both FPGA and ASIC platforms.
The high-level design remains identical for both targets. The only
significant low-level difference lies in the implementation of the
core 32$\times$32-bit multiplier: the FPGA version leverages the platform's
dedicated DSP blocks, while the ASIC version employs a custom-designed
multiplier built from standard cells. This unified architecture is
structured hierarchically to optimize for two critical metrics across
both targets: resource efficiency (e.g., LUT/DSP usage for FPGA, gate
count for ASIC) and maximizing clock speed.
Fig.~\ref{fig:arch} illustrates the high-level architecture of the proposed CSIDH accelerator.

\subsection{ALU Implementations}

Our ALU comprises the following specialized
low-level operators:
\begin{itemize}
\item \textit{Adder:} Designed for 512-bit modular addition.
\item \textit{Subtractor:} Implements 512-bit modular subtraction.
\item \textit{Multiplier:} Performs 512-bit by 512-bit modular multiplication within
the finite field.
\end{itemize}
\subsubsection{Adder} 
\begin{figure}
\includegraphics[bb=110bp 360bp 550bp 600bp,clip,width=1\columnwidth]{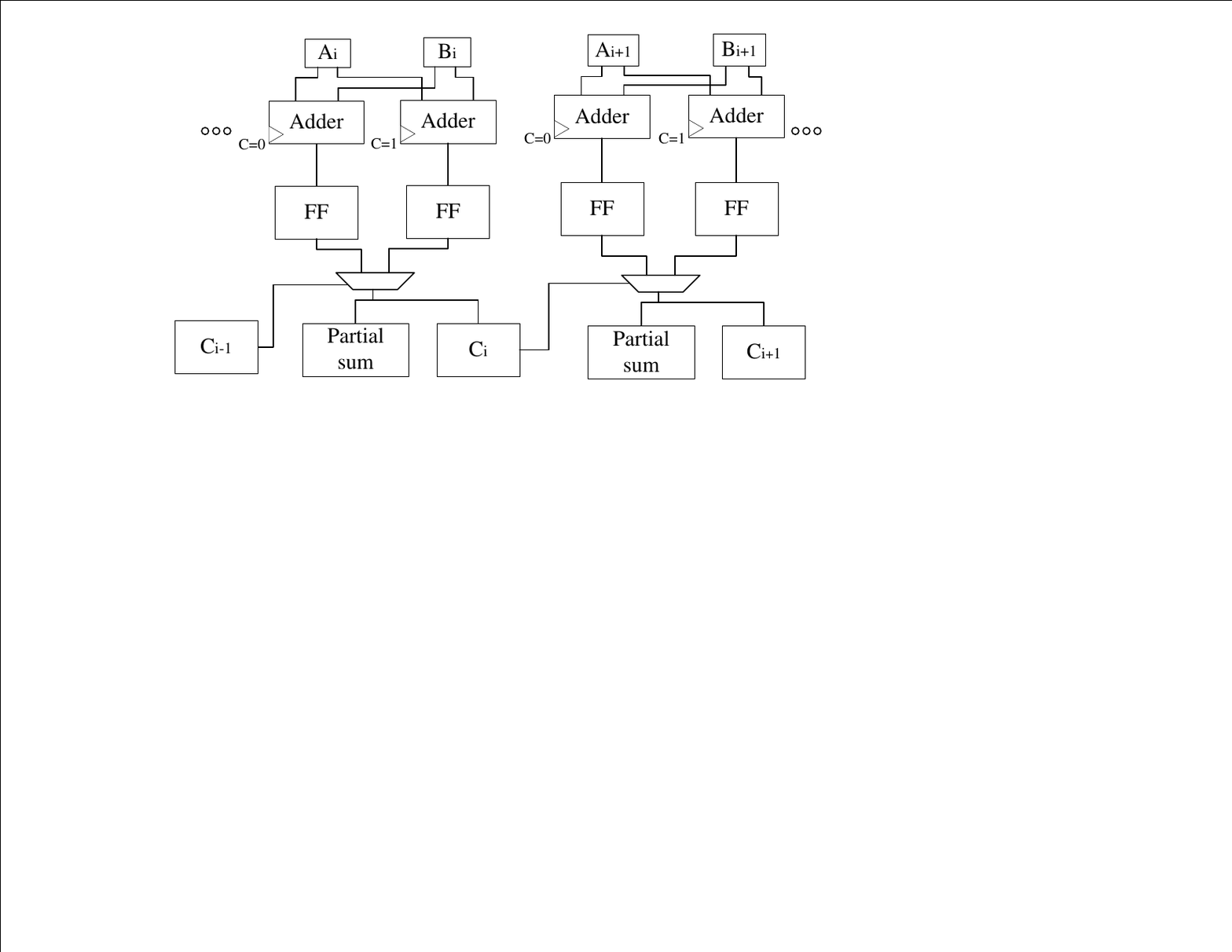}\caption{Architecture of the pipelined carry-select adder. To minimize critical
path delay, each 32-bit chunk is processed in parallel by pre-calculating
sums for both a potential carry-in of 0 and 1. A multiplexer then
selects the correct precomputed result based on the actual carry
propagated from the previous stage.}
\label{fig:adder}
\end{figure}

To handle large integers efficiently, our adder and subtractor
operate on 32-bit words. In initial timing experiments, a monolithic
512-bit adder running at 100 MHz exhibited a negative slack of 26
ns. Prior work shows that a 32-bit fast carry-chain can reliably clock
at up to 400 MHz, on par with 32-bit multipliers, so
we adopted a 32-bit word size for our design \cite{mongomery2}.
Our proposed adder is implemented with a technique similar to that
proposed in \cite{work in addtion}, partitioning each 512-bit operand
into sixteen 32-bit chunks, as illustrated in Fig.~\ref{fig:adder}. For each chunk, two sums are computed
in parallel:
\begin{itemize}
\item \textit{Scenario1:} Addition with carry-in = 0.
\item \textit{Scenario 2:} Addition with carry-in = 1.
\end{itemize}
A multiplexer, driven by the carry-out of the previous chunk, selects
the correct sum and carry-out for the current chunk. To minimize critical-path
delay, we pipeline this into two stages:
\begin{itemize}
\item \textit{Stage1:} Compute and register both partial sums (for carry-in = 0 and
carry-in = 1).
\item \textit{Stage2:} Propagate the actual carry and select the correct registered
sum.
\end{itemize}

By confining the longest combinational logic to a single 32-bit adder,
our design reduces the critical path to just one fast 32-bit addition, typically
only a few nanoseconds on modern FPGAs. The procedure for the adder is presented in Algorithm \ref{alg-adder}.

\subsubsection{Subtractor} The subtractor employs a similar chunk-based approach,
replacing carry with borrow. For each 32-bit chunk, two differences
are computed: 
\begin{itemize}
\item \textit{Scenario1:} Subtraction with borrow-in = 0.
\item \textit{Scenario 2:} Subtraction with borrow-in = 1.
\end{itemize}
The borrow-out from the previous chunk selects the correct difference
and borrow-out, maintaining a critical path comparable to the adder.
This design ensures both operations are fast and resource-efficient,
critical for CSIDH\textquoteright s frequent arithmetic computations.

\begin{algorithm}[t]
$\textbf{Input:}$$a,b\:\in\digamma_{p}$, $word=32$, $chunk$= 16

$\textbf{Output:}$$S\:\in\digamma_{p}$, $C_{out}$

---------------------------------------------------------------------------

// Stage 1: Dual-Path Sum Generation

1: $\text{\textbf{for}}$ $i\leftarrow0\:to\:chunk-1$ \textbf{do}

2:\textbf{ $\quad$}$(s_{0}[i],c_{0}[i])\leftarrow Add(a[i],b[i],0)$

3:\textbf{ $\quad$}$(s_{1}[i],c_{1}[i])\leftarrow Add(a[i],b[i],1)$

4: \textbf{end for}

// Stage 2: Carry Propagation and Sum Selection

5: $\text{\textbf{for}}$ $i\leftarrow0\:to\:chunk-1$ \textbf{do}

6:\textbf{ $\quad$}$S[i]\leftarrow mux(c[i-1],s_{1}[i],s_{0}[i])$

7:\textbf{ $\quad$}$c[i]\leftarrow mux(c[i-1],c_{1}[i],c_{0}[i])$

8: \textbf{end for}

9: $S\leftarrow concat(S[chunk-1],...,S[0])$

10: $C_{out}=c[chunk-1]$

\caption{\foreignlanguage{american}{512-bit carry-select adder (two-stage pipeline).}}
\label{alg-adder}
\raggedright{}\vspace{1mm}
\end{algorithm}
\subsubsection{Multiplier} CSIDH algorithms heavily rely on multiplication within
a finite field of characteristic p. Therefore, to address the significant
challenge of division overhead when multiplying two 512-bit numbers
and obtaining a 512-bit result, we utilize the Montgomery modular
multiplication. The core idea behind Montgomery modular multiplication
is to transform numbers into a \textquotedbl Montgomery domain\textquotedbl{}
where modular multiplication can be computed more efficiently. This
involves an auxiliary modulus $R$, typically chosen as a power of 2
(like $2^{512}$ in the provided steps) for computational ease. To
multiply two numbers $a$ and $b$ modulo $p$, one first converts them to
their Montgomery representations ($a*R\pmod p$ and $b*R\pmod p$).
The standard multiplication of these representations yields $a*b*R^{2}\pmod p)$.
The Montgomery reduction then efficiently divides this product by
$R$ by performing a 512-bit right shift to obtain $a*b*R\pmod p$, which
is the Montgomery form of the desired result. A final conversion out
of the Montgomery domain (by another Montgomery reduction with 1)
is needed if the standard representation $a*b\pmod p$ is required.
However, in the CSIDH algorithm, all intermediate multiplications
remain in the Montgomery domain, eliminating the need for conversions
until the very last step, where the final result is converted back
to standard form.

\begin{algorithm}[t]
$\textbf{Input:}$ $T$ (1024-bit integer, where $T=a*b$ for 512-bit
inputs a, b)

$\textbf{Input:}$ $p$ (512-bit prime modulus)

$\textbf{Input:}$ $R=2^{512}$ (constant)

$\textbf{Input:}$ $pinv=-p^{-1}\pmod R$ (precomputed)

$\textbf{Output:}$ $T_{out}\equiv T*R^{-1}\pmod p$ (512-bit result
in $[0,p-1]$)

\textemdash\textemdash\textemdash\textemdash\textemdash\textemdash\textemdash\textemdash\textemdash\textemdash\textemdash\textemdash\textemdash\textemdash\textemdash\textemdash\textemdash\textemdash\textemdash\textemdash\textemdash\textemdash\textemdash\textemdash\textemdash{}

1: \textbf{Compute} $T_{low}=T\pmod R$ (lower 512 bits of $T$)

2: \textbf{Compute} $m=(T_{low}*pinv)\pmod R$

3: \textbf{Compute} $T'=T+m*p$ (1024-bit addition) 

4: \textbf{Compute} $T_{out}=T'/R$ (Right shift $T'$by 512 bits)

5: \textbf{if $T_{out}>p$}

6:\textbf{\quad{}$T_{out}=T_{out}-p$}

7: \textbf{return} $T_{out}$

\caption{The Montgomery Reduction Algorithm \cite{mont_main}}
\label{algo2}
\raggedright{}\vspace{1mm}
\end{algorithm}
The Montgomery reduction algorithm, as detailed above and illustrated in Algorithm \ref{algo2}, necessitates
efficient 512$\times$512-bit multiplication for computing intermediate products
such as $T=a*b$. To achieve this within an FPGA environment, we evaluated
several multiplication strategies, ultimately optimizing for maximal
utilization of DSP blocks and minimal critical path delays.

We considered various approaches, starting with Karatsuba multiplication.
This method theoretically reduces multiplication complexity from $O(n^{2})$
to approximately $O(n^{1.585})$ using a divide-and-conquer strategy
\cite{Main CSIDH Article}. However, for 512-bit operands, our analysis indicated that
while it requires roughly 81 32-bit multiplications compared to 256
in the traditional schoolbook method, it introduces significant recursive
overhead. This includes increased addition and subtraction operations,
along with complex control logic for managing intermediate results,
making it less efficient for our target FPGA implementation due to
DSP block underutilization and potential routing delays.
Other advanced methods like Toom-Cook, a generalization of Karatsuba
requiring even more operand splitting and recombination \cite{jalali work on constant time}, and
FFT-based multiplication, which is generally efficient for operands
larger than 1024 bits but incurs substantial pre and post processing
overhead for 512-bit numbers \cite{main constant time}, were also considered unsuitable
for our specific requirements.

In contrast, the schoolbook multiplication method, while having a
baseline complexity of $O(n^{2}$), offers significant potential for
parallelization and efficient utilization of FPGA DSP blocks, especially
when the operands are divided into smaller chunks like 32-bit segments.
This aligns well with modern FPGAs, whose dedicated DSP blocks can
perform a 32$\times$32-bit multiplication in just \textasciitilde 5 nanoseconds.
This reliance on pre-optimized hardware, however, creates a significant
design challenge for an ASIC implementation, which lacks such specialized
blocks. Our analysis with industry-standard synthesis tools indicates
that a single-cycle 32$\times$32-bit multiplier built from standard cells
requires a clock period of at least 7.5 nanoseconds, posing a substantial
challenge to achieving high clock speeds. Nevertheless, as we will
explain, our pipelined architecture is specifically designed to overcome
this bottleneck, enabling the ASIC to reach a higher frequency than
the FPGA without increasing the total clock cycle count.

\textbf{Architecture:} Our multiplication strategy is based on processing one
32-bit chunk of operand $A$ with all 16 32-bit chunks of operand
$B$ per clock cycle. This approach is systematically divided into
three main phases:
\begin{enumerate}
\item \textit{Partial Product Generation:} In each cycle, the $i^{th}$ 32-bit chunk
of $A$, denoted as $A_{i}$, is multiplied with all 16 chunks of
$B$, $B_{j}$ for $j=0\:to\:15$. This results in 16 partial products,
each 64 bits wide. These intermediate values are subsequently stored
in registers for the next processing stage. For our FPGA implementation,
this entire phase completes in a single clock cycle by leveraging
the device's dedicated DSP blocks.
\begin{equation}
P_{ij}=A_{i}*B_{j\:}for\:j=0\:to\,15.
\end{equation}
\item \textit{Intermediate Summation of Partial Products:} The subsequent challenge
involves efficiently summing the 16 partial products to generate a
single 17$\times$32-bit wide result. As illustrated in Fig.~\ref{fig:adder}, due
to the alignment of 32-bit chunks, there exists a 32-bit overlap between
successive products. This overlap facilitates the application of the
previously discussed carry-select addition strategy. Specifically,
two scenarios for each sum (carry-in = 0 and carry-in = 1) are computed
and stored in registers. This phase requires two clock cycles to produce
the final result of the intermediate summation of partial products.
\begin{equation}
Pi=\sum_{j=0}^{15}P_{ij}.
\end{equation}
\item \textit{Accumulation with the Final Result Register:} The accumulated sum from
the previous step is then added to a 1024-bit result register, an
operation that completes in two clock cycles. A critical data dependency
exists at this stage, as the calculation for the current chunk $A_{i}$
cannot begin until the final accumulated value from the previous chunk
$A_{i-1}$ is available. This dependency introduces a necessary one-cycle
stall in the pipeline between processing each chunk to ensure the
correct result.
\begin{equation}
result=\sum_{i=0}^{15}P_{i}.
\end{equation}
\end{enumerate}

Our architecture transforms this apparent bottleneck into a key design
advantage. This built-in stall provides the timing budget needed to
relax the performance constraints on the initial partial product generation
stage. Consequently, for our ASIC implementation, we designed the
32$\times$32-bit multipliers as two-cycle units, the specific architecture
of which will be detailed further. This design easily meets the timing
requirements for a 180 MHz clock frequency, whereas a single-cycle
implementation would have created a critical path limiting the speed.
The extra cycle required for multiplication is almost entirely absorbed
by the pipeline stall, resulting in only a minimal one-cycle increase
to the total latency of the full 512$\times$512-bit multiplication, while
enabling a higher overall operating frequency.

After 16 iterations, we obtain the full 512$\times$512-bit multiplication
result. This baseline implementation completes the multiplication
in 36 clock cycles. At a clock frequency of 200 MHz, this translates
to a total latency of approximately 180 ns.
Given the performance-critical nature of CSIDH, which involves many
large integer multiplications and its inherent timing weaknesses,
we further optimized the multiplier by exploiting parallelism across
operand chunks to increase speed. Specifically, the 16 chunks of A
are split into two batches:
\begin{itemize}
\item \textit{Upper Batch:} Chunks $A_{8}$ to $A_{15}$ 
\item \textit{Lower Batch:} Chunks $A_{0}$ to $A_{7}$ 
\end{itemize}
Each batch is processed in parallel using the same three-phase multiplication
and summation pipeline, supported by separate accumulation registers
for the upper and lower products. After both parallel pipelines are
completed, a final summation is performed between the upper and lower
accumulation registers to produce the final 512$\times$512-bit result.

\begin{figure}
\includegraphics[bb=107bp 349bp 638bp 590bp,clip,width=1\columnwidth]{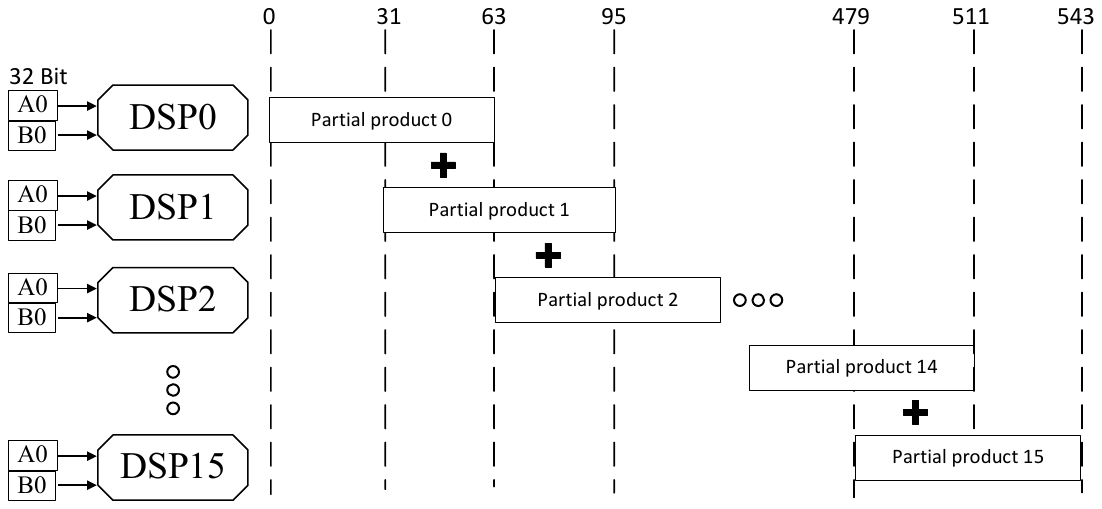}\caption{Architecture of the parallelized 512$\times$512-bit multiplier. In a single
cycle, one 32-bit chunk of operand A is multiplied against all 16
chunks of operand B using an array of DSPs to generate 16 partial
products simultaneously. These products, staggered according to their
bit-weight, are then efficiently summed by a pipelined adder tree.}
\label{fig:multiplier}
\end{figure}

This enhanced, parallel architecture reduces the baseline multiplication
time from 36 cycles. For the FPGA implementation, the total latency
is 22 clock cycles, corresponding to 110 ns at 200 MHz. For the ASIC
implementation, the latency is just one cycle longer at 23 clock cycles,
the ASIC achieves a latency of 126.5 ns at 180 MHz. As illustrated in Fig.~\ref{fig:multiplier}, our parallelized architecture processes one 32-bit chunk of operand~\textit{A} in a single cycle by multiplying it against all 16 chunks of operand~\textit{B} using an array of DSPs to generate 16 partial products simultaneously.The complete flow of this parallel and pipelined multiplication process is formally presented in Algorithm \ref{alg-mul}.

\begin{algorithm}[t]
$\textbf{Input:}$Two 512-bit integers $a,b\:\in\digamma_{p}$

\textbf{Input: }$word=32$, $chunk=16$ // 512 bits = 16 words of
32 bits

$\textbf{Output:}$$P=a\times b$

---------------------------------------------------------------------------

// Phase 0: Initialize up/down accumulators

1: $ACC_{\uparrow}\leftarrow0^{1024}$, $ACC_{\downarrow}\leftarrow0^{1024}$

// Phase 1 (per-chunk partials): multiply one 32-bit chunk of \ensuremath{a}
by all 16 chunks of \ensuremath{b}

2: $\text{\textbf{for}}$ $k\leftarrow0\:to\:(chunk/2)-1$ \textbf{do}

3:\foreignlanguage{american}{$\quad$}$\textbf{for}$ $i\leftarrow0\:to\:chunk-1$
\textbf{do}

4:\textbf{ $\quad$}$\quad$$(lo[i],hi[i])\leftarrow Mul32(a[k], b[i]))$
// 64-bit partial: $lo[i]=ppi[31:0],hi[i]=ppi[63:32]$

5: $\quad$ \textbf{end for}

$\quad$// Phase 2 (overlap folding with carry-select): build the 544-bit chunk product

6: $\quad$$(W[0],c[0])\leftarrow Add32CS(lo[0],0,0)$

7:\foreignlanguage{american}{ $\quad$}$\textbf{for}$ $t\leftarrow0\:to\:chunk$
\textbf{do}

8:\textbf{ $\quad$}$\quad$$(W[t],c[t])\leftarrow Add32CS(hi[t-1],lo[t],c[t-1])$

9: $\quad$ \textbf{end for}

$\quad$// Phase 3: accumulate into 1024-bit vector at the proper
word offset

12:\textbf{ $\quad$$CHUNK\leftarrow concat(W[chunk],\ldots,W[0])$
// }17\texttimes 32 = 544 bits

13:\textbf{ $\quad$$ACC_{\uparrow}\leftarrow ACC_{\uparrow}+(CHUNK\ll(32\times k))$}

14: \textbf{end for}

15: $\textbf{for}$ $v\leftarrow(chunk/2)\:to\:chunk-1$ \textbf{do}

16: $\quad$repeat Steps 3--13 with $a[v]$ and accumulate into $ACC_{\downarrow}$
at shift $32\times v$

17: \textbf{end for}

// Phase 4: Final merge of up/down vectors

18: $P\leftarrow ACC\uparrow+ACC_{\downarrow}$

19: \textbf{return }$P$

\caption{\foreignlanguage{american}{512\texttimes 512-bit Multiplier}}
\label{alg-mul}
\raggedright{}\vspace{1mm}
\end{algorithm}
\textit{Two-Cycle 32$\times$32-bit Multiplier Design:} Our ASIC 32$\times$32-bit multiplier
is optimized for high-frequency operation with a fixed latency of
two clock cycles. To achieve this, the design is centered on the Radix-4
Booth's algorithm, which is advantageous as it reduces the number
of partial products that must be summed from 32 to just 17. As the
Booth algorithm is inherently designed for signed operands, we handle
our unsigned multiplication by padding each 32-bit input with a '0'
at the most significant bit (MSB). This effectively transforms the
operation into a 33$\times$33-bit signed multiplication where both operands
are positive, yielding a correct 64-bit unsigned result without requiring
complex correction logic.

The multiplication is partitioned into two balanced pipeline stages
to meet the high-frequency target. In the first stage, the 33-bit
multiplier operand is Booth-encoded to generate the 17 partial products.
A combinational adder tree then reduces the first nine of these products
into an intermediate sum, which is latched into a pipeline register.
The remaining eight partial products are also latched concurrently.
In the second stage, a separate adder tree sums the intermediate value
from the first stage with the remaining eight partial products to
produce the final 64-bit result. By splitting the adder tree across
two stages, we create two combinational paths of roughly equivalent
delay, which is critical for maximizing the clock frequency.

This two-stage pipelined Booth architecture was chosen after considering
standard alternatives. A simple Array multiplier, while regular in
layout, is unsuitable for high-performance applications due to its
long critical path delay. Conversely, fast combinational multipliers
like Wallace and Dadda trees offer high speed for single-cycle operations
but introduce significant layout complexity and routing challenges
due to their irregular interconnect structure. Our proposed architecture
strikes an optimal balance: it reduces the partial product count significantly
and uses a balanced pipeline to achieve a high clock frequency with
a minimal two-cycle latency, making it an excellent choice for the
computationally intensive operations in CSIDH \cite{mul32,mul32_2}.

\subsection{Control Unit Implementation }

The control unit acts as the central coordinator for the CSIDH protocol's
cryptographic operations. It efficiently directs the shared ALU to execute the necessary field arithmetic, effectively
managing the flow of Algorithm \ref{algo1}. This control logic is implemented
as a hierarchy of FSMs, where higher-level
FSMs invoke lower-level ones to manage the sequence of operations.
This modular and reusable design provides precise control over the
protocol's complex computations while maximizing ALU utilization,
a key feature introduced in Section 3.1. The control unit comprises
five key FSM-driven modules: xDBLADD, xAffinize, xMul, xISOG,
xPRNG, and the top-level CSIDH module, each tailored to
a specific task within the CSIDH key-exchange process. These key modules
are described as follows:

\subsubsection{xISOG} The xISOG module is the most computationally intensive
FSM in our design, responsible for executing a complete isogeny computation
of a given odd degree $l$. This FSM takes as input the curve coefficient
$A$, a point $P$ on the curve, and the kernel point $K$ that defines
the isogeny. Upon completion, it overwrites $A$ and $P$ with their new
values on the target curve. The FSM orchestrates the ALU to implement
a highly optimized algorithm that proceeds in three main phases:
\begin{itemize}
\item \textit{Initialization:} The process begins by initializing two distinct running
products. The first is for evaluating the isogeny's effect on the
point $P$. The second is for calculating the coefficient $A'$ of
the new curve, which involves a temporary representation in Edwards
coordinates.
\item \textit{Core Computation Loop:} The FSM enters a loop that iterates $(l-1)/2$
times. In each iteration, it performs two main tasks in sequence:
\begin{itemize}
\item \textit{Kernel Point Generation:} It computes  the next multiple of the kernel
point, $[i]K$, via a highly efficient differential addition chain, reusing results to reduce field multiplications and using a sliding register window to minimize memory.
\item \textit{Product Accumulation:} As each new multiple of $K$ is generated, it is
immediately \textquotedbl consumed\textquotedbl{} into the two running
products initialized in the first phase. This updates both the point
evaluation product and the curve coefficient product.
\end{itemize}
\item \textit{Finalization and Fault Detection:} After the loop completes, the FSM
finalizes the results. The first accumulated product is used to compute
the final coordinates of the new point, $\varphi(P)$. The second
product is used in a formula that converts the curve to and from the
Edwards representation to efficiently calculate the new curve coefficient,
$A'$. Additionally, the module supports an optional fault detection
mechanism. When enabled, it computes $[l]K$ at the end of the process.
The FSM verifies that this result is the point at infinity (Z-coordinate
is zero); any other result indicates a computational fault, ensuring
the integrity of the operation.
\end{itemize}
\subsubsection{xAffinize} Responsible for converting projective coordinates
to affine coordinates, this module computes the modular inverse of
the Z-coordinate using a single Montgomery inversion. The resulting
affine x-coordinate is critical for the final key derivation step
of the CSIDH protocol.

\subsubsection{xMul} This module executes scalar multiplication of an elliptic
curve point $P$ by an integer $k$. It sequences point doubling
and addition operations, utilizing the xDBLADD module to minimize
redundant computations and optimize performance.

\subsubsection{xDBLADD} This module simultaneously performs point doubling
and point addition on Montgomery-projective coordinates. By leveraging
shared intermediate values, it reduces the number of field multiplications
by half compared to performing these operations separately \cite{Main CSIDH Article}.
This optimization enhances computational efficiency, making it a cornerstone
of the design.

\subsubsection{xTWIST} This module determines whether a given x-coordinate
belongs to a point on the primary Montgomery curve or its quadratic
twist. It functions by evaluating the curve equation's right-hand
side and performing a quadratic residue test on the result.

\subsubsection{CSIDH} The top-level module orchestrates the entire key-exchange
process. It integrates the lower-level modules to perform the group
action that generates a public key from a private key. To ensure the
integrity of the final output, our accelerator includes an optional
validation step for the computed curve coefficient, $A^\prime$. This
procedure performs a sanity check to verify that the resulting Montgomery
curve, $E_{A^\prime}:y^{2}=x^{3}+A^\prime x^{2}+x$,
is a valid supersingular curve over $\digamma_{p}$.The validation
is performed by first generating a random point P on the curve $E_{A\prime}$.
Then, we compute the scalar multiplication $[p+1]P$. For any valid
supersingular curve over our chosen prime field, the group of $\digamma_{p}$-rational
points has order $p+1$. Therefore, by Lagrange's theorem, the result
of this operation must be the point at infinity $(\mathcal{O})$.
Observing this outcome confirms that the coefficient $A'$ corresponds
to a curve with the correct group order, providing confidence that
a catastrophic computational error has not occurred. While this method
validates the basic structure of the resulting curve, it is not a
formal proof of correctness against all possible fault-injection attacks.

\section{Security Analysis} \label{sec-sec}

The CSIDH-512 cryptographic scheme was initially estimated to offer
approximately 128-bit classical security. However, its security remains
an active area of research, relying on unproven assumptions concerning
the computational hardness of isogeny problems. Furthermore, practical
implementations of CSIDH face significant side-channel attack risks
\cite{security1}. A critical vulnerability arises because the CSIDH
secret key directly influences the number of small-degree isogenies
computed in each loop iteration. Consequently, both computation time
and power consumption become key-dependent. As observed by Meyer et
al. \cite{optimize constant time2}, a worst-case key (e.g., all exponents
= +5) can lead to execution times more than three times longer than
an average key, while an all-zero key results in minimal computation.
This variability implies that an unprotected CSIDH implementation
inherently leaks secret key information through its runtime characteristics.
Similarly, an attacker capable of measuring power or electromagnetic
(EM) traces can distinguish between the algorithm's two primary operations
(field multiplication vs. isogeny steps) and isolate individual loop
iterations. In such a model, the attacker can deduce the prime factors
present in the key and even their signs, drastically reducing the
key search space \cite{optimize constant time2}.

These timing and power side-channel vulnerabilities underscore the
imperative for constant-time design. In response, several research
groups have developed fixed-time CSIDH algorithms. Meyer et al. \cite{optimize constant time2}
introduced the concept of \textquotedbl dummy isogenies,\textquotedbl{}
ensuring a fixed number of isogenies of each degree are always computed,
thus standardizing the workload irrespective of the secret key. Subsequent
works by Onuki et al. \cite{faster isogeny2} and Cervantes et al.
\cite{security2} built upon this concept. Notably, Cervantes et al. \cite{security2}
identified and addressed subtle oversights in prior constant-time
proposals, even introducing a dummy-free fixed-time CSIDH algorithm
(leveraging Edwards-curve arithmetic) that incurs only a constant
factor overhead \cite{security2}. More recently, Jalali et al. \cite{jalali work on constant time}
provided the first comprehensive constant-time CSIDH implementation
on 64-bit ARM, designed to resist timing attacks by consistently performing
the full operation. Furthermore, Banegas et al. \cite{banegas} introduced
CTIDH, a novel key space and group-action algorithm that nearly doubles
the speed of constant-time CSIDH-512, reducing the isogeny count by
approximately half, from \textasciitilde 200 million to \textasciitilde 125.5
million clock cycles.

\subsection{Constant time CSIDH}

To mitigate the side-channel vulnerabilities described above, our
proposed FPGA architecture implements a constant-time version of the
CSIDH group action based on the strategy from Campos, F. et al.\cite{main constant time}. The
core of our approach is to ensure that the sequence of operations
is independent of the private key. This is achieved by using two exponent
vectors: the true private key, $e$, and a \textquotedbl constant-time\textquotedbl{}
vector, $e_{ct}$, which is derived from $e$. For each index $i$,
the value $e_{ct}[i]$ is set to the maximum absolute value of the
exponent range (e.g., +5 or -5, depending on the sign of $e[i]$).

The main control FSM of our hardware always executes the isogeny chain
corresponding to the $e_{ct}$ vector, thus performing a fixed, maximum
number of computations for every key. Within each step of the computation,
the FSM checks the value of the true exponent $e[i]$. This check
determines whether to perform a \textquotedbl real\textquotedbl{}
or a \textquotedbl dummy\textquotedbl{} isogeny:
\begin{itemize}
\item \textit{Real Isogeny:} If $e[i]$ is non-zero, the operation proceeds normally.
The newly computed curve coefficient $A'$ and the projected point
$\varphi(P$) are stored, updating the state for the next step. The
corresponding value in $e[i]$ is then decremented or incremented. As
part of this process, a verification step is performed: the kernel
point $K$ is multiplied by its order $l_{i}$. Since $K$ is a point
of order $l_{i}$, the result must be the point at infinity (where
the Z-coordinate is zero). This provides an online fault detection
mechanism within each step, as a non-infinity result would indicate
a computational error and invalidate the output.
\item \textit{Dummy Isogeny:} If $e[i]$ is zero, the hardware still performs the
full, computationally expensive isogeny calculation. However, the
results ($A'$ and $\varphi(P)$) are conditionally discarded, and
the state registers are not updated. This ensures the workload and
power profile remain identical to a real operation. A crucial step
in the dummy operation is handling the point $P$ for the next isogeny
in the batch. Since the projected point $\varphi(P)$ is discarded,
the point $P$ is instead updated by multiplying it by the current
isogeny's degree, $l_{i}$. This effectively removes the $l_{i}$-torsion
from $P$, yielding a valid point on the original curve to be used
in the next iteration.
\end{itemize}

This constant-time logic was implemented in our hardware's control
unit and was also integrated into the C reference code from \cite{Main CSIDH Article}
to create a customized software benchmark. This ensures that our FPGA
accelerator's performance is fairly evaluated against an equivalently
secure software implementation. Ultimately, this strategy ensures
that the computational workload is identical for both real and dummy
operations. 

\subsection{Masked ALU}

To provide a robust defense against power analysis attacks, our architecture
moves beyond a constant-time FSM and incorporates masking directly
at the hardware's computational core: the ALU. The countermeasure
is designed to make different ALU instructions, such as 'ADD' versus
'MULTIPLY', indistinguishable from a power consumption perspective.

In a conventional design, executing a specific command activates only
the necessary hardware. For example, an 'ADD' command would engage
the adder submodule, while the multiplier would remain idle, creating
a unique and recognizable power signature for that operation. An adversary
could exploit these command-specific signatures to reverse-engineer
the high-level algorithm.

Our ALU design eradicates this vulnerability at the submodule level.
We ensure that all major arithmetic units, the adder, subtractor,
and multiplier, are computationally active during every
clock cycle, regardless of which command is being executed. This is
managed by the ALU's internal control logic:

When a specific command is issued, such as 'MULTIPLY', the multiplier
submodule receives the real operands from the main data path to perform
the genuine computation. Simultaneously, the adder and subtractor
submodules, which are not needed for this command, are fed with randomly
generated numbers to perform \textquotedbl dummy\textquotedbl{} operations
whose results are discarded. Conversely, during an 'ADD' command,
the adder performs the real operation while the multiplier and subtractor
execute dummy operations with random inputs.

This strategy of parallel real and dummy computations ensures that
the overall switching activity within the ALU remains high and statistically
uniform across all supported commands. The power signature of an addition
becomes nearly identical to that of a multiplication, effectively
hiding the nature of the instruction being processed from a side-channel
adversary. This powerful security guarantee is a deliberate design
trade-off, accepted in exchange for higher overall power consumption,
to achieve a robust defense against sophisticated power analysis attacks.

\section{Implementation Results} \label{sec-results}

The proposed CSIDH hardware accelerator is synthesized and implemented
for two distinct hardware targets: a Xilinx Zynq UltraScale+ FPGA
and a 180nm ASIC. The FPGA implementation of the 512-bit design requires
approximately 66.3k LUTs, 47k flip-flops, and 128 DSPs, while the
1024-bit design utilizes approximately 140.1k LUTs, 91.8k flip-flops,
and 256 DSPs. A full breakdown of the FPGA resources by module is
provided in Table \ref{Table I}.

\begin{figure}
\includegraphics[bb=72bp 253bp 540bp 720bp,clip,width=1\columnwidth]{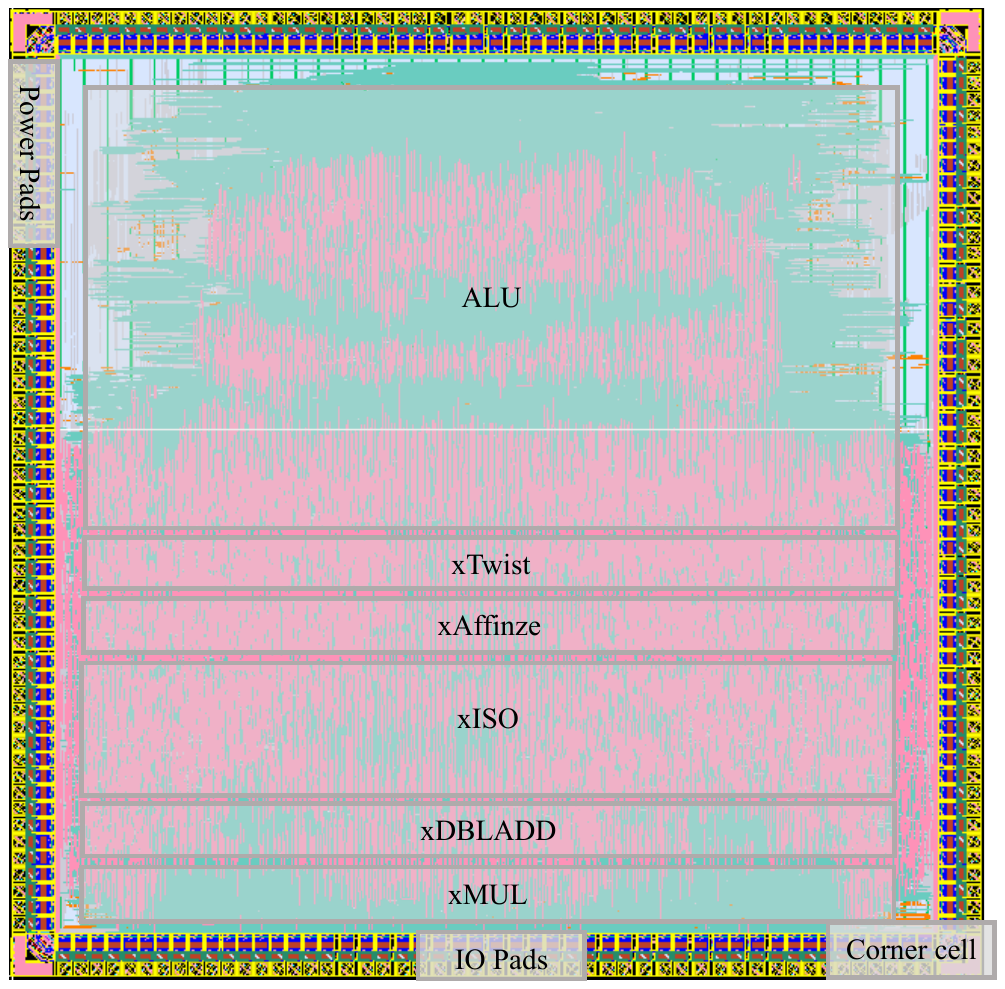}\caption{Final floorplan of the 512-bit CSIDH accelerator IC. The chip occupies
an area of 4.1$\times$4.1 mm$^{2}$, which includes 196 signal I/O pads and
4 power pads. The approximate placement of the primary functional
units is shown.}
\label{fig:floorplan}
\end{figure}

For the ASIC target, the design is implemented through a full place-and-route
flow using a 180nm process. The final layout includes a 200-pad ring,
comprising 196 signal pads and 4 dedicated power pads (two for the
core logic and two for the I/O ring), as shown in Fig. \ref{fig:floorplan}. The data pads are configured
as bidirectional to serve as inputs or outputs depending on the accelerator's
operational state. For the 512-bit configuration, the final chip has
dimensions of 4.1$\times$4.1 mm and contains 566,223 standard cell instances.
To gauge the scalability of the architecture, the 1024-bit configuration
was synthesized, but the full place-and-route flow was not completed
for this larger design. Post-synthesis estimates indicate that the
1024-bit version requires 856,592 cell instances and would occupy
an estimated area of 5.6$\times$5.6 mm$^{2}$. A detailed breakdown of the
effective resource utilization for both the final 512-bit implementation
and the estimated 1024-bit design is presented in Table \ref{Table II}.

\begin{table}[t]
\caption{Post-implementation resource utilization by module for the 512-bit
and 1024-bit CSIDH designs on a Xilinx Zynq UltraScale+ FPGA}
\label{Table I}
\centering{}\centering%
\begin{tabular}{llcccccc}
\toprule 
\multicolumn{2}{l}{Module} & \multicolumn{2}{c}{LUT} & \multicolumn{2}{c}{FF} & \multicolumn{2}{c}{DSP}\tabularnewline
\midrule 
\multicolumn{2}{l}{Setting} & 512 & 1024 & 512 & 1024 & 512 & 1024\tabularnewline
\midrule
\multicolumn{2}{l}{ALU} & 24207 & 54762 & 0 & 0 & 128 & 256\tabularnewline
\multirow{6}{*}{CU} & xMUL & 2505 & 5726 & 0 & 0 & 0 & 0\tabularnewline
 & xDBLADD & 5021 & 9999 & 0 & 0 & 0 & 0\tabularnewline
 & xAffinize  & 6819 & 13080 & 0 & 0 & 0 & 0\tabularnewline
 & xTwist & 3153 & 6936 & 0 & 0 & 0 & 0\tabularnewline
 & xISO & 18684 & 39249 & 0 & 0 & 0 & 0\tabularnewline
 & CSIDH & 5917 & 9833 & 46954 & 91875 & 0 & 0\tabularnewline
\midrule 
\multicolumn{2}{l}{Total} & 66306 & 139585 & 46954 & 91875 & 128 & 256\tabularnewline
\bottomrule
\end{tabular}
\end{table}

\begin{table}[t]
\caption{Post-Place and Route area and Gate utilization results for the 180nm ASIC implementation}
\label{Table II}
\centering{}\centering%
\begin{tabular}{llcccc}
\toprule 
\multicolumn{2}{l}{Module} & \multicolumn{2}{c}{Gate Count} & \multicolumn{2}{c}{Area ($mm^{2}$)}\tabularnewline
\midrule 
\multicolumn{2}{l}{Setting} & 512 & 1024 & 512 & 1024\tabularnewline
\midrule
\multicolumn{2}{l}{ALU} & 386655 & 522352 & 7.941 & 11.61\tabularnewline
\multirow{6}{*}{CU} & xMUL & 14330 & 24011 & 0.370 & 0.575\tabularnewline
 & xDBLADD & 20176 & 39386 & 0.527 & 1.039\tabularnewline
 & xAffinize & 15668 & 31714 & 0.371 & 0.745\tabularnewline
 & xTwist & 6168 & 12104 & 0.153 & 0.304\tabularnewline
 & xISO & 51920 & 102324 & 1.421 & 2.779\tabularnewline
 & CSIDH & 71306 & 124701 & 1.972 & 3.838\tabularnewline
\midrule 
\multicolumn{2}{l}{Total} & 566223 & 856592 & 12.755 & 20.89\tabularnewline
\bottomrule
\end{tabular}
\end{table}

\begin{table*}[t]
\label{table2}\caption{Performance and resource utilization comparison of the proposed modular
multiplier with other hardware implementations}
\label{Table III}
\centering{}\centering%
\begin{tabular}{llllllcccccc}
\toprule 
\multicolumn{6}{l}{Design (Ref)} & FPGA & Bit & LUT & DSP & Clock (MHz) & Latency (ns)\tabularnewline
\midrule
\multicolumn{6}{l}{Our Design} & Virtex Ultrascale+ & 512 & 24207 & 128 & 200 & 435\tabularnewline
\multicolumn{6}{l}{Ozturk, Erdinc, et al. \cite{mongomery1}} & Virtex Ultrascale+ & 512 & 329868 & 1089 & 38 & 25.68\tabularnewline
\multicolumn{6}{l}{Gary C.T. Chow, et al. \cite{mongomery2}} & Virtex-6 & 512 & 62557 & 324 & 300 & 10\tabularnewline
\multicolumn{6}{l}{Amiet, Dorian, et al. \cite{Mongomery3}} & Virtex-7 & 512 & 1917 & 9 & 225 & 5310\tabularnewline
\multicolumn{6}{l}{San, Ismail, et al. \cite{Mongomery3}} & Virtex Ultrascale+ & 512 & 84735 & 2592 & 88.8 & 33.75\tabularnewline
\multicolumn{6}{l}{Zhang, Yuxuan, et al. \cite{mongomery5}} & \multirow{1}{*}{Virtex-7} & 1024 & 10588 & 0 & \multirow{1}{*}{370} & \multirow{1}{*}{720}\tabularnewline
\multicolumn{6}{l}{Our Design} & Virtex Ultrascale+ & 1024 & 54762 & 258 & 200 & 675\tabularnewline
\bottomrule
\end{tabular}
\end{table*}

\begin{table*}[t]
\label{table3}\caption{End-to-end performance comparison of the proposed FPGA and ASIC accelerator
with existing CSIDH software implementations}
\label{Table IV}
\centering{}\centering%
\begin{tabular}{llllllccccc}
\toprule 
\multicolumn{6}{l}{Design (Ref)} & Platform & Bit & Clk cycles (M) & Clock Freq. (MHz) & Latency\tabularnewline
\midrule
\multicolumn{6}{l}{Castryck, W. et al. \cite{Main CSIDH Article}} & Intel Core i7-7700 & 512 & \textasciitilde 1222 & 3000 & \textasciitilde 407 ms\tabularnewline
\multicolumn{6}{l}{Campos, F. et al. \cite{main constant time}} & ARM Cortex-M4 & 512 & 15,523 & 24 & 646 s\tabularnewline
\multicolumn{6}{l}{Jalali, A. et al. \cite{jalali work on constant time}} & Cortex-A57 & 512 & 11,256 & 1,950 & 5.79 s\tabularnewline
\multicolumn{6}{l}{Castryck, W. et al. \cite{Main CSIDH Article}} & Intel Core i7-7700 & 1024 & \textasciitilde 5550 & 3000 & \textasciitilde 1.850 s\tabularnewline
\midrule 
\multicolumn{6}{l}{Our Design (FPGA)} & Virtex Ultrascale+ & 512 & 103 & 200 & 515 ms\tabularnewline
\multicolumn{6}{l}{Our Design (FPGA)} & Virtex Ultrascale+ & 1024 & 482 & 100 & 4.821 s\tabularnewline
\multicolumn{6}{l}{Our Design (ASIC)} & SMIC 180nm CMOS & 512 & 106 & 180 & 591 ms\tabularnewline
\multicolumn{6}{l}{Our Design (ASIC)} & SMIC 180nm CMOS & 1024 & 496 & 60 & \textasciitilde 8.27 s\tabularnewline
\bottomrule
\end{tabular}
\end{table*}

\subsection{ALU Performance}

In this section, we provide a comparative analysis of our proposed ALU against several notable hardware implementations from the literature, a summary of which is presented in Table \ref{Table III}.
Work such as \cite{mongomery1,mongomery2,mongomery4},
prioritize performance, often at the expense of substantial hardware
area. These implementations typically consume a considerable amount
of FPGA resources, which may not be practical for CSIDH. In contrast,
other research, exemplified by \cite{Mongomery3}, aims to minimize
resource consumption, which often leads to poor performance.
The latter approach is generally suboptimal for CSIDH, given its
stringent timing requirements. Our design philosophy necessitates
a pragmatic trade-off between performance and area. Although the parameters
presented in \cite{mongomery5} appear promising in terms of this
balance, a critical consideration for CSIDH is the requirement for
a classic (non-modular) multiplier and modular adder and subtractor.
The architecture proposed in \cite{mongomery5}, along with similar
works, does not inherently support classic multiplication and does
not have modular subtractor and adder for the CSIDH modular operation.
Therefore, our proposed ALU offers a compelling solution, capable
of efficiently handling both classic and Montgomery modular multiplication
and having a modular subtractor and adder, thereby addressing the
diverse computational demands of CSIDH.

\subsection{CSIDH Performance Analysis and Comparison}

As noted in the introduction, this work presents, to our knowledge,
the first comprehensive hardware implementation of CSIDH on both FPGA
and ASIC platforms. To provide a clear overview of our accelerator's
performance, we compare it against existing software and microcontroller-based
implementations, with a full comparison detailed in Table \ref{Table IV}.

Leveraging our proposed architecture, a constant-time CSIDH-512 key
generation on our FPGA target requires $1.03\times10^{8}$ clock cycles,
resulting in a latency of 515 ms at 200 MHz. The ASIC implementation
requires slightly more cycles at $1.065\times10^{8}$, at 180 MHz
clock speed yields a faster latency of 591 ms. This small increase
in cycle count is a direct consequence of our two-cycle 32$\times$32-bit
multiplier, which adds one additional cycle per multiplication operation.
The challenges of accelerating larger parameter sets are highlighted
by the CSIDH-1024 configuration. As shown in the resource breakdown
in Tables \ref{Table I} and \ref{Table II}, the hardware cost for the 1024-bit design is approximately
double that of the 512-bit version, revealing a significant scalability
challenge for this architecture. 

For the FPGA implementation, the
design's complexity and routing congestion limited the final operating
frequency to 100 MHz, resulting in a key generation latency of 4.82
seconds (from $4.82\times10^{8}$cycles). For the ASIC version of
the 1024-bit design, preliminary post-synthesis results provide a
frequency estimate of 60 MHz. For the required $4.96\times10^{8}$
cycles, this corresponds to an estimated latency of 8.27 seconds.
It should be noted that this is a conservative estimate from initial
synthesis, which could be improved with further place-and-route optimization.
However, both the final FPGA result and the preliminary ASIC estimate
highlight the significant increase in hardware complexity and timing
closure challenges when scaling the CSIDH architecture to larger security
parameters.

For comparison, software implementations on general-purpose processors
typically demand significantly higher cycle counts. For example, the
main CSIDH code in constant-time mode, executed on an Intel Core i7-7700
core, requires approximately $1.222\times10^{9}$ cycles (approximately
382 ms at 3.2 GHz) for a single constant-time group-action computation.
The 1024-bit software equivalent takes approximately $5.55\times10^{9}$
clock cycles, completing a single constant-time group-action computation
in nearly 2 seconds \cite{Main CSIDH Article}.

The performance gap is even more pronounced on embedded microcontrollers.
A constant-time implementation of CSIDH on a Cortex-A57 processor,
as reported in Jalali, A. et al. \cite{jalali work on constant time},
requires $11,286\times10^{6}$ clock cycles. At an operating frequency
of 1.95 GHz, this translates to approximately 5.79 seconds for a constant-time
group action. On more constrained embedded microcontrollers, such
as the ARM Cortex-M4, the disparity is even larger. The work presented
in Campos, F. et al. \cite{main constant time} reports a requirement
of $15,523\times10^{6}$ clock cycles on an STM32F407. At a clock
rate of 24 MHz, this corresponds to approximately 10.78 minutes to
complete a constant-time group action. In stark contrast, our FPGA
design completes the same operation in well under one second.

\begin{table*}[t]
\label{table4}\caption{Performance and resource comparison of the proposed CSIDH accelerator
with other PQC KEM hardware implementations on FPGA}
\label{Table V}
\centering{}\centering%
\begin{tabular}{llllllccccccc}
\toprule 
\multicolumn{6}{l}{Design (Ref)} & FPGA & Public\nobreakdash-Key Size (B) & LUT & DSP & KeyGen Latency & Clock (MHz) & Reference\tabularnewline
\midrule
\multicolumn{6}{l}{CSIDH\textendash 512 } & Virtex Ultrascale+ & 64 & 66306 & 128 & 515 ms & 200 & This work\tabularnewline
\multicolumn{6}{l}{CSIDH\textendash 1024 } & Virtex Ultrascale+ & 128  & 139585 & 256 & 4.821 s & 100 & This work\tabularnewline
\multicolumn{6}{l}{Kyber-512} & Artix-7 & 800 & 14000 & 2 & \textasciitilde 5.3 $\mu$s & 208 & Ni, Z. et al. \cite{kyber}\tabularnewline
\multicolumn{6}{l}{Kyber-1024} & Artix-7 & 1184 & 15700 & 2 & \textasciitilde 13 $\mu$s & 208 & Ni, Z. et al. \cite{kyber}\tabularnewline
\multicolumn{6}{l}{Saber-768} & Virtex Ultrascale+ & 1088 & 23700 & 0 & \textasciitilde 21.8 $\mu$s & 250 & Roy, S. et al. \cite{Saber}\tabularnewline
\multicolumn{6}{l}{NTRU-Prime} & Virtex Ultrascale+ & 930 & 9538 & 19 & \textasciitilde 4.808 ms & 271 & Marotzke, A. et al. \cite{NTSR}\tabularnewline
\multicolumn{6}{l}{SIKE-p751} & \multirow{1}{*}{Virtex-7} & 564 & 90900 & 834 & \textasciitilde 9.27 ms & \multirow{1}{*}{370} & \multirow{1}{*}{Tian, J. et al. \cite{SIKE}}\tabularnewline
\bottomrule
\end{tabular}
\end{table*}

\subsection{Performance Comparison with PQC Hardware Implementations}

To evaluate the standing of our work within the broader post-quantum
landscape, Table \ref{Table V} provides a comparative analysis of our CSIDH FPGA
accelerator against state-of-the-art hardware implementations of other
PQC Key Encapsulation Mechanisms (KEMs). This benchmark facilitates
a direct comparison of the critical trade-offs between performance,
resource utilization, and public key size among lattice-based KEMs
(Kyber, Saber, NTRU-Prime) and isogeny-based schemes.

As shown in Table \ref{Table V}, FPGA implementations of lattice-based KEMs are
characterized by extremely high performance and efficient resource
utilization. The NIST primary standard, CRYSTALS-Kyber, achieves key
generation latencies in the microsecond range; for instance, Kyber-512
completes its operation in approximately 5.3 \textmu s using only
14,000 LUTs on an Artix-7 FPGA \cite{kyber}. Similarly, Saber-768
operates at a comparable speed of \textasciitilde 21.8 \textmu s
on a Virtex Ultrascale+ device \cite{Saber}. Even NTRU-Prime delivers
a key generation time of \textasciitilde 4.8 ms, which is two orders
of magnitude faster than our CSIDH implementation \cite{NTSR}. In
stark contrast, isogeny-based cryptosystems present a different profile
of high resource demand and significantly higher latency. For comparison,
we include the high-performance implementation of SIKE-p751. It is
crucial to note that the SIKE algorithm was cryptographically broken
in 2022 \cite{sike_break}. However, its hardware implementation remains
an valuable benchmark as it was a prominent isogeny-based candidate,
and its architecture represents the significant resource costs associated
with high-security isogeny computations. The SIKE-p751 implementation
is faster than our CSIDH design at \textasciitilde 9.27 ms, but this
performance comes at a resource cost, consuming over 90,000 LUTs and
834 DSPs \cite{SIKE}.

The data clearly illustrates that a significant performance gap, spanning
several orders of magnitude, exists between the hardware capabilities
of lattice-based and isogeny-based schemes. However, this performance
disparity must be weighed against other critical metrics, most notably
public key size, where CSIDH holds a compelling advantage. The compactness
of CSIDH, with a public key over 12 times smaller than that of Kyber,
makes it a highly attractive candidate for applications where bandwidth
or storage is severely constrained, such as in certain Internet of Things (IoT) or embedded
systems. The primary challenge, which this work addresses, is whether
this extreme performance cost can be mitigated through hardware acceleration
to make such a trade-off viable.

\section{Conclusion and future work} \label{sec-conclusion}

In this work, we have introduced a comprehensive
hardware accelerator for the CSIDH key-exchange protocol, presenting
a unified architecture for both FPGA and ASIC platforms supporting
512-bit and 1024-bit configurations. This development establishes
a crucial hardware performance baseline on both reconfigurable and
custom silicon for isogeny-based cryptography, a field that, despite
its promising security attributes, has remained relatively underexplored
in hardware. 
The impact of this work extends beyond key exchange, as the hardware
acceleration principles are directly applicable to the entire family
of isogeny-based protocols. This is particularly relevant for emerging
digital signature schemes like SQISign, which is an active candidate
in the ongoing NIST process to standardize additional post-quantum
signatures. Given the high computational demands of such schemes,
the development of efficient hardware accelerators, as detailed in
this paper, will be essential for their practical deployment.

Our results demonstrate that constant-time CSIDH-512 key generation
can be achieved in 515 ms on our target FPGA and accelerated further
to 591 ms on a 180nm ASIC. We believe that this work opens avenues for further research.
Future efforts could focus on algorithmic refinements or circuit-level
improvements to the architecture presented. Now that a baseline has
been established on a mature process, a clear path forward involves
targeting more advanced ASIC nodes. On a modern 3nm process, for example,
it is estimated that the operating frequency could scale into the
low-gigahertz range, potentially reducing key generation times to
well under 100ms. This would make our architecture ideal for integration
as a compact and efficient accelerator core within a larger System-on-Chip
(SoC) for secure IoT applications. We hope these initial results encourage
the community to develop more sophisticated hardware for isogeny-based
cryptosystems, diversifying the post-quantum ecosystem with robust,
high-performance options.

\end{document}